\documentclass[10pt,twocolumn,american,english,aps,reprint,prl,amsmath]{revtex4}
\usepackage[T1]{fontenc}
\usepackage[latin9]{inputenc}
\usepackage{amsmath}
\usepackage{graphicx}

\makeatletter

\@ifundefined{textcolor}{}
{%
 \definecolor{BLACK}{gray}{0}
 \definecolor{WHITE}{gray}{1}
 \definecolor{RED}{rgb}{1,0,0}
 \definecolor{GREEN}{rgb}{0,1,0}
 \definecolor{BLUE}{rgb}{0,0,1}
 \definecolor{CYAN}{cmyk}{1,0,0,0}
 \definecolor{MAGENTA}{cmyk}{0,1,0,0}
 \definecolor{YELLOW}{cmyk}{0,0,1,0}
}

\usepackage{subfigure}
\usepackage{hyperref}

\usepackage{breqn}
\gdef\wrap@breqn@environ#1#2{
    \expandafter\let\csname breqn@oldbegin@#1\expandafter\endcsname\csname #1\endcsname
    \expandafter\let\csname breqn@oldend@#1\expandafter\endcsname\csname end#1\endcsname
    \expandafter\gdef\csname breqn@begin@#1\endcsname{%
        \expandafter\let\csname #1\expandafter\endcsname\csname breqn@oldbegin@#1\endcsname%
        \begin{#2}%
    }
    \expandafter\gdef\csname breqn@end@#1\endcsname{%
        \expandafter\let\csname end#1\expandafter\endcsname\csname breqn@oldend@#1\endcsname%
        \end{#2}%
        \expandafter\let\csname #1\expandafter\endcsname\csname breqn@begin@#1\endcsname%
        \expandafter\let\csname end#1\expandafter\endcsname\csname breqn@end@#1\endcsname%
    }
    \expandafter\let\csname #1\expandafter\endcsname\csname breqn@begin@#1\endcsname
    \expandafter\let\csname end#1\expandafter\endcsname\csname breqn@end@#1\endcsname
}
\wrap@breqn@environ{equation}{dmath}
\wrap@breqn@environ{equation*}{dmath*}
\DeclareRobustCommand\[{\begin{equation*}}
\DeclareRobustCommand\]{\end{equation*}}

\makeatother

\usepackage{babel}
\begin{document}

\title{Near the jamming transition of elastic active cells: A sharp-interface
approach}

\author{Yony Bresler}

\author{Benoit Palmieri}

\author{Martin Grant}
\email{martin.grant@mcgill.ca}

\affiliation{Physics Department McGill University Montreal Canada}

\date{\today}
\begin{abstract}
We use a sharp interface model for active cells to study the jamming
transition point and behavior near it by varying cell concentration,
active velocity and elasticity, including a binary mixture of soft
and stiff cells. We determine the jamming transition point, as well
as behavior near the transition, including the effective diffusion,
and sixfold bond correlations. Finally, we expand on previous studies
by showing the Voronoi dimensionless cell shape can be treated as
an order parameter at any concentration.
\begin{description}
\item [{Usage}] Secondary publications and information retrieval purposes.{\small \par}
\item [{PACS~numbers}] May be entered using the environment \textsf{PACS~numbers}.{\small \par}
\end{description}
\end{abstract}

\pacs{64.70.pm, 87.18.Fx}

\keywords{Cell jamming, Cell unjamming, Collective cellular migration, cell
shape, glass transition}

\maketitle
Assemblies of cells can behave as a liquid, or as a solid. Recently,
the study of how biological processes relate to concepts such as jamming
and the glass transition have gained attention. These studies include
experiments, such as jamming during cancer invasion \cite{Haeger2014a,Oswald2017},
ongoing jamming in the corneal endothelium \cite{Brookes2017}, self-driven
jamming in microbial growth \cite{Delarue2016}, and unjamming in
wounds \cite{Chepizhko2018}. There have also been theoretical biophysics
studies, which have established a connection between cell migration
and glass dynamics \cite{Angelini2011}, direct measurement of the
transition with varying concentration for soft disks \cite{Henkes2011},
and a proposed jamming phase diagram \cite{Sadati2013}. Notably,
the Vertex \cite{Bi2015} and Self-propelled Voronoi \cite{Bi2016}
models showed that cell shape can be used to determine the jamming
state. These studies have also recently been extended to 3D \cite{Nogucci2018}. 

Though driven and active system are far from equilibrium, they have
been shown to share key features with the equilibrium glass transition,
such as effective thermal behavior and time correlation functions
\cite{Berthier2009,Berthier2013}. However, despite a wide range of
experimental and theoretical work on the glass transition, several
fundamental questions remain unanswered or contested \cite{Langer2014,Sillescu1999,Parisi2010}.
In two dimensions, one must consider a potential Kosterlitz-Thouless-Halperin-Nelson-Young
hexatic phase \cite{Kosterlitz1973,Steinhardt1983}. Indeed, this
has been seen, for example in colloidal films \cite{Peng2010} and
in repulsive Monte Carlo simulations \cite{Prestipino2011}, although
for hard-core or very short-range interactions, transitions appear
to be first-order \cite{Bernard2011}. Whatever the case may be for
active systems, these results for thermal systems will play a role.

In this letter we wish to extend these studies using our recently
developed model for elastic cells \cite{Bresler2018}, a sharp interface
limit of a phase field model of cells \cite{Palmieri2015}, which
shares some similarities with the recently released \textit{CellSim3D}
model \cite{Madhikar2018}. Points along a cell interface $\mathbf{R}\left(\theta\right)$,
with elasticity $\gamma$ evolve by minimizing the local curvature
$K$, setting a preferred cell area $\pi R_{0}^{2}$, in the presence
of repulsion forces due to the other cells, and the cell velocity.
More precisely, the time evolution for cell $n$ is given in dimensionless
units by

\begin{equation}
\partial_{t}\mathbf{R}_{n}\left(\theta,t\right)=\left[\gamma K\left(\theta,t\right)-\frac{1}{R_{0}}+\mu\left(A_{n}-\pi R_{0}^{2}\right)+\beta\left(\tanh\left(\alpha d_{n,\theta}\right)-1\right)^{2}\right]\mathbf{n}_{\theta}+\mathbf{v}_{n},\label{eq:drdt total}
\end{equation}
 where $K\left(\theta,t\right)$ is the local curvature of the interface,
$\frac{1}{R_{0}}$ is the natural curvature of a spherical cell, $\mu$
sets the strength of the preferred cell area, $A_{n}$ is the current
area, $\lambda$ is the interface thickness, $\alpha=\sqrt{\frac{15}{2}}/\lambda$,
$\beta=\frac{150}{\alpha\lambda}$ and $d_{n,\theta}$ is the distance
to the nearest neighboring cell along the local normal to the interface
$\mathbf{n}_{\theta}$. The cell velocity $\mathbf{v}_{n}$ is the
sum of active and inactive terms: The active velocity has a constant
magnitude $v_{A}$, and reorients with probability $P(t)=\frac{1}{\tau}e^{-t/\tau}$,
where $\tau$ is the mean time between reorientations. Inactive velocity
is due to forces exerted by the other cells surrounding it as well
as the substrate and surrounding water. 
\begin{figure*}
\selectlanguage{american}%
\centering%
\makebox[0.3\paperwidth]{%
\subfigure[]{\label{fig:signatures-of-Jamming:snapshot}\foreignlanguage{english}{\includegraphics[bb=420bp 55bp 1060bp 700bp,clip,width=0.2\textwidth]{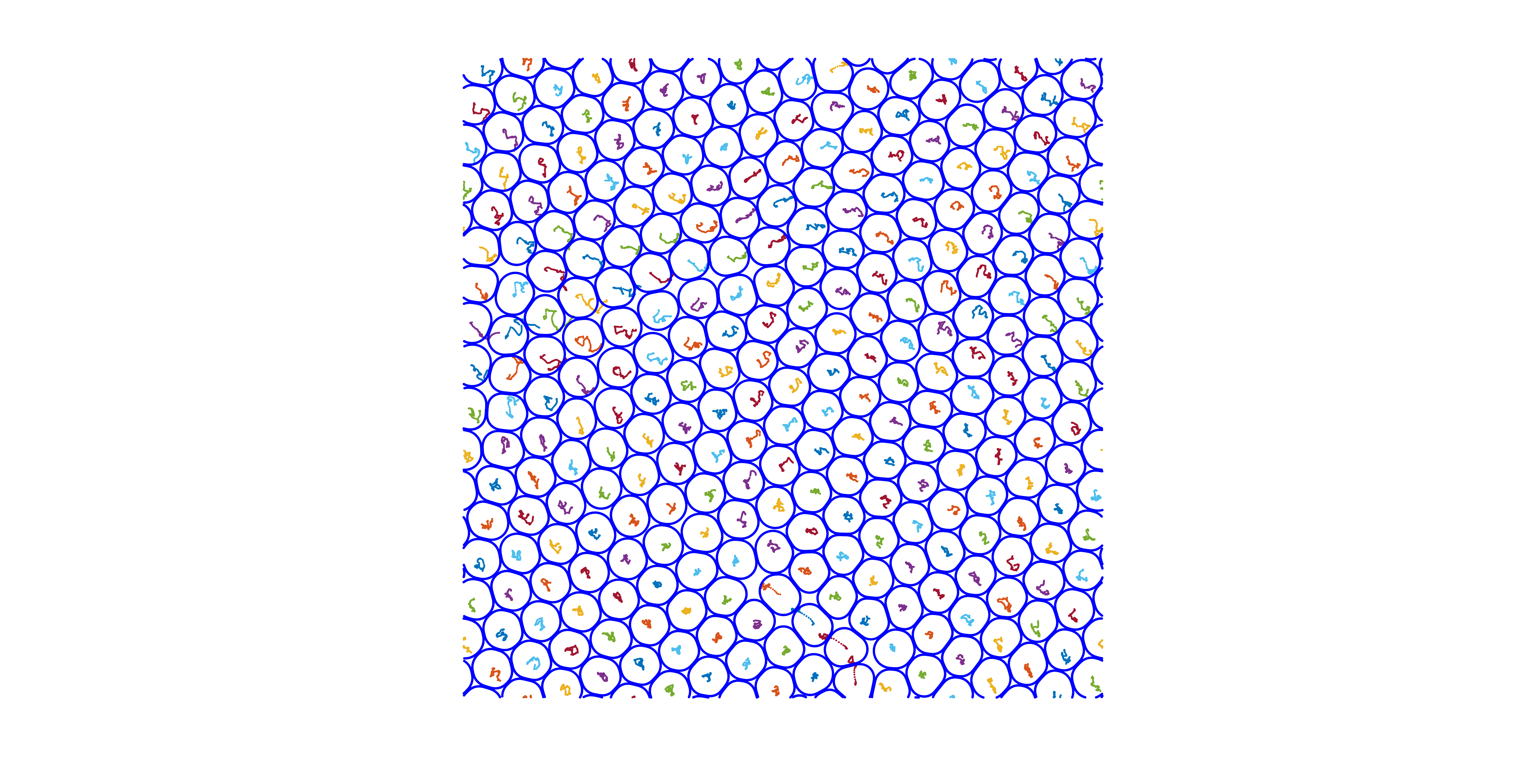}}}%
}\hspace{2.5in}\centering%
\makebox[0.3\paperwidth]{%
\subfigure[]{\label{fig:signatures-of-Jamming:MSD}\foreignlanguage{english}{\includegraphics[bb=0bp 0bp 1350bp 785bp,width=0.3\textwidth]{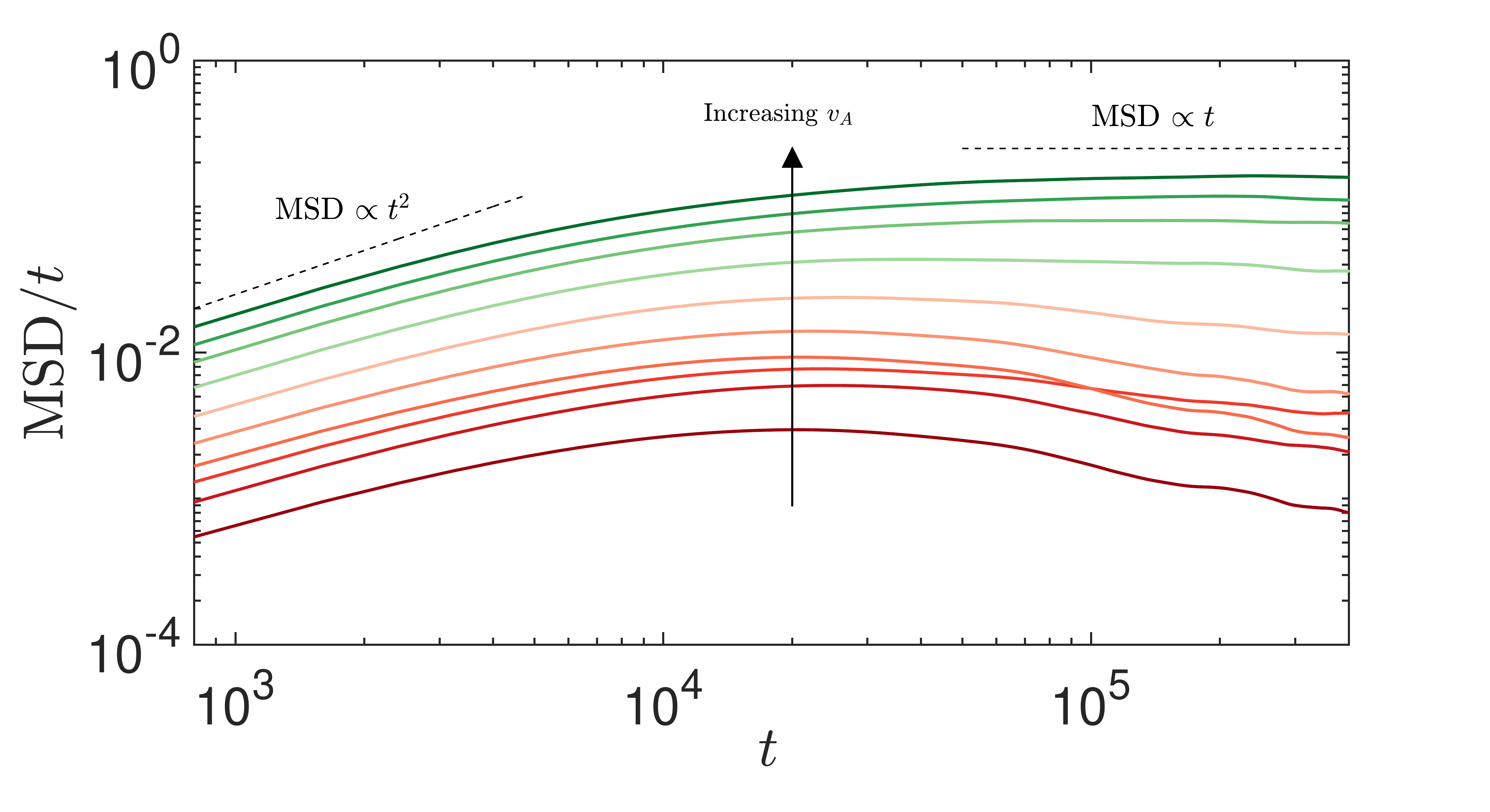}}}%
}\hspace{2.5in}\centering%
\makebox[0.3\paperheight]{%
\subfigure[]{\label{fig:signatures-of-Jamming:g6}\foreignlanguage{english}{\includegraphics[bb=30bp 0bp 1340bp 695bp,width=0.3\textwidth]{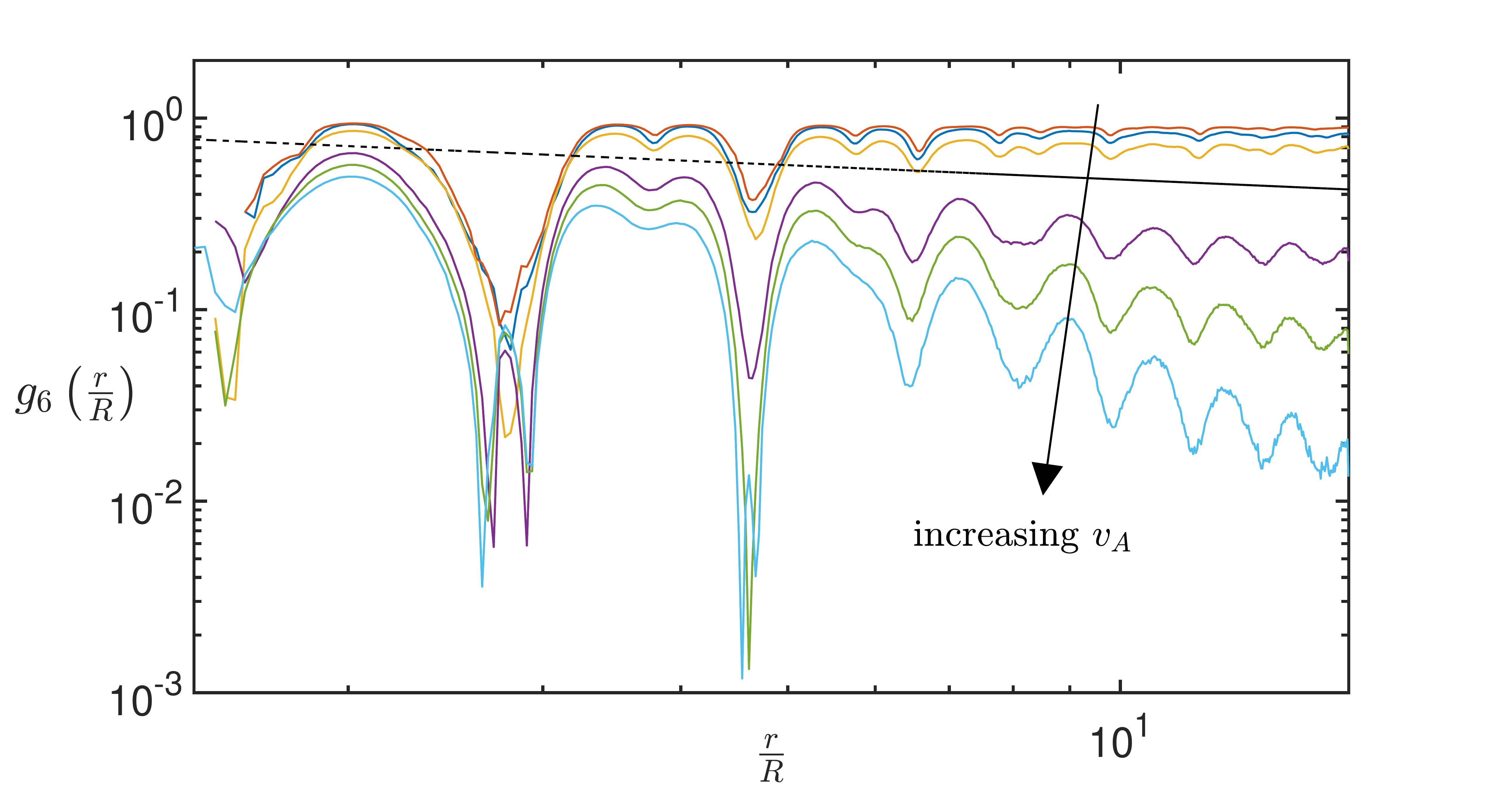}}}%
}\\
\centering%
\makebox[0.25\paperwidth]{%
\subfigure[]{\label{fig:signatures-of-Jamming:D_vs_v}\foreignlanguage{english}{\includegraphics[width=0.3\textwidth]{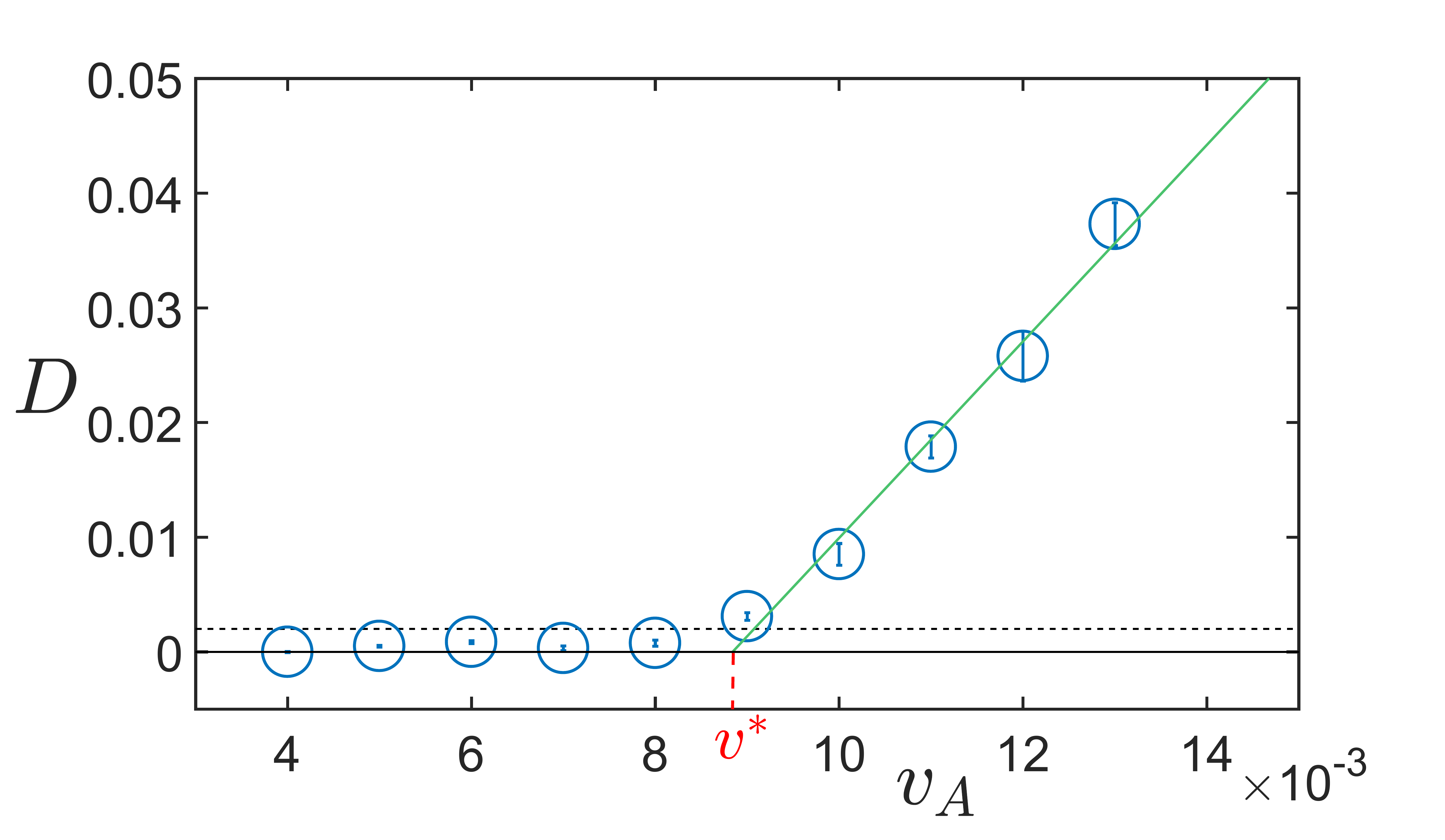}}}%
}\hspace{2.5in}\centering%
\makebox[0.25\paperwidth]{%
\subfigure[]{\label{fig:signatures-of-Jamming:D_vs_rho}\foreignlanguage{english}{\includegraphics[width=0.3\textwidth]{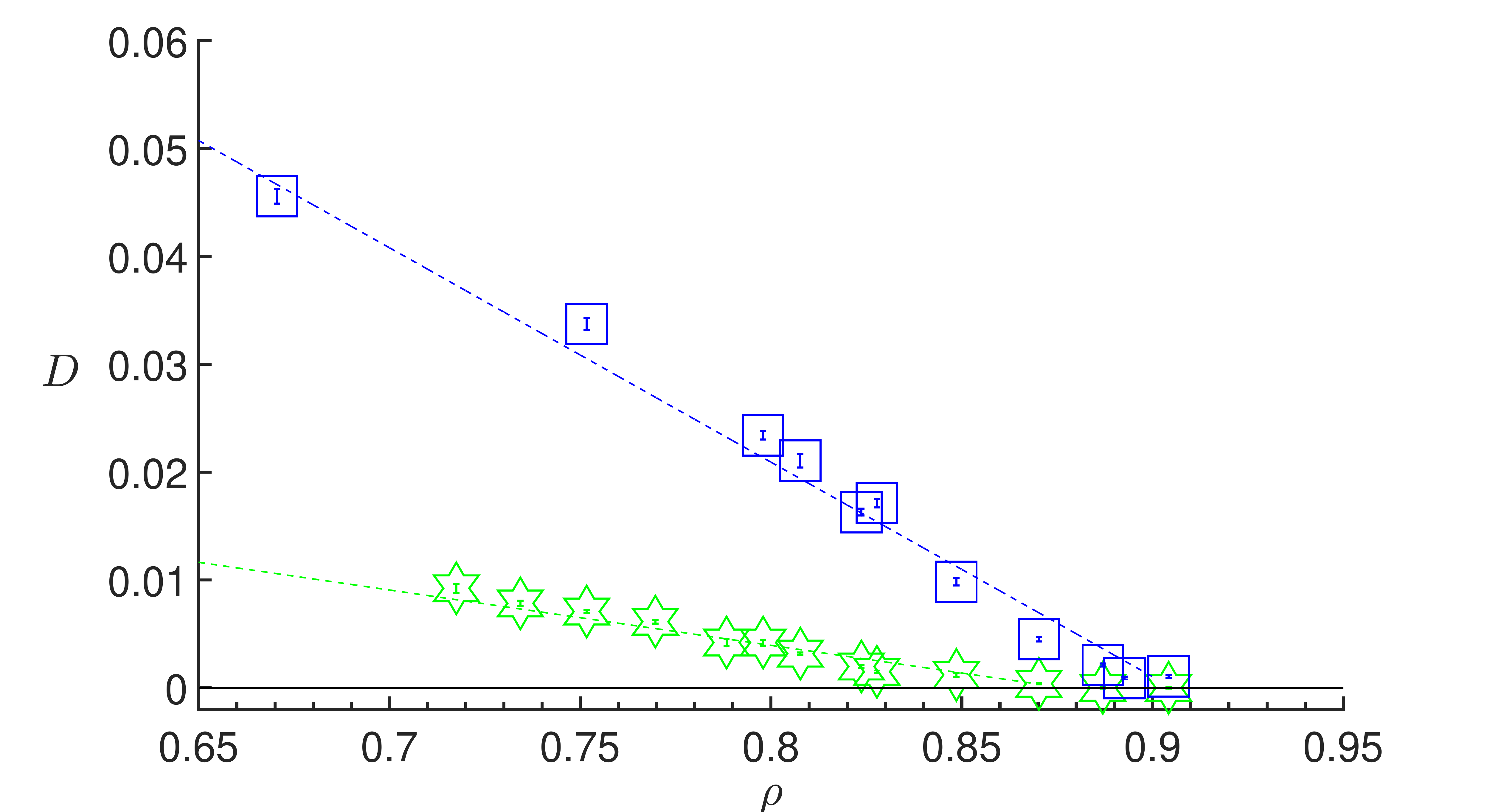}}}%
}\hspace{2.5in}\centering%
\makebox[0.25\paperwidth]{%
\subfigure[]{\label{fig:signatures-of-Jamming:v*_vs_rho}\foreignlanguage{english}{\includegraphics[width=0.3\textwidth]{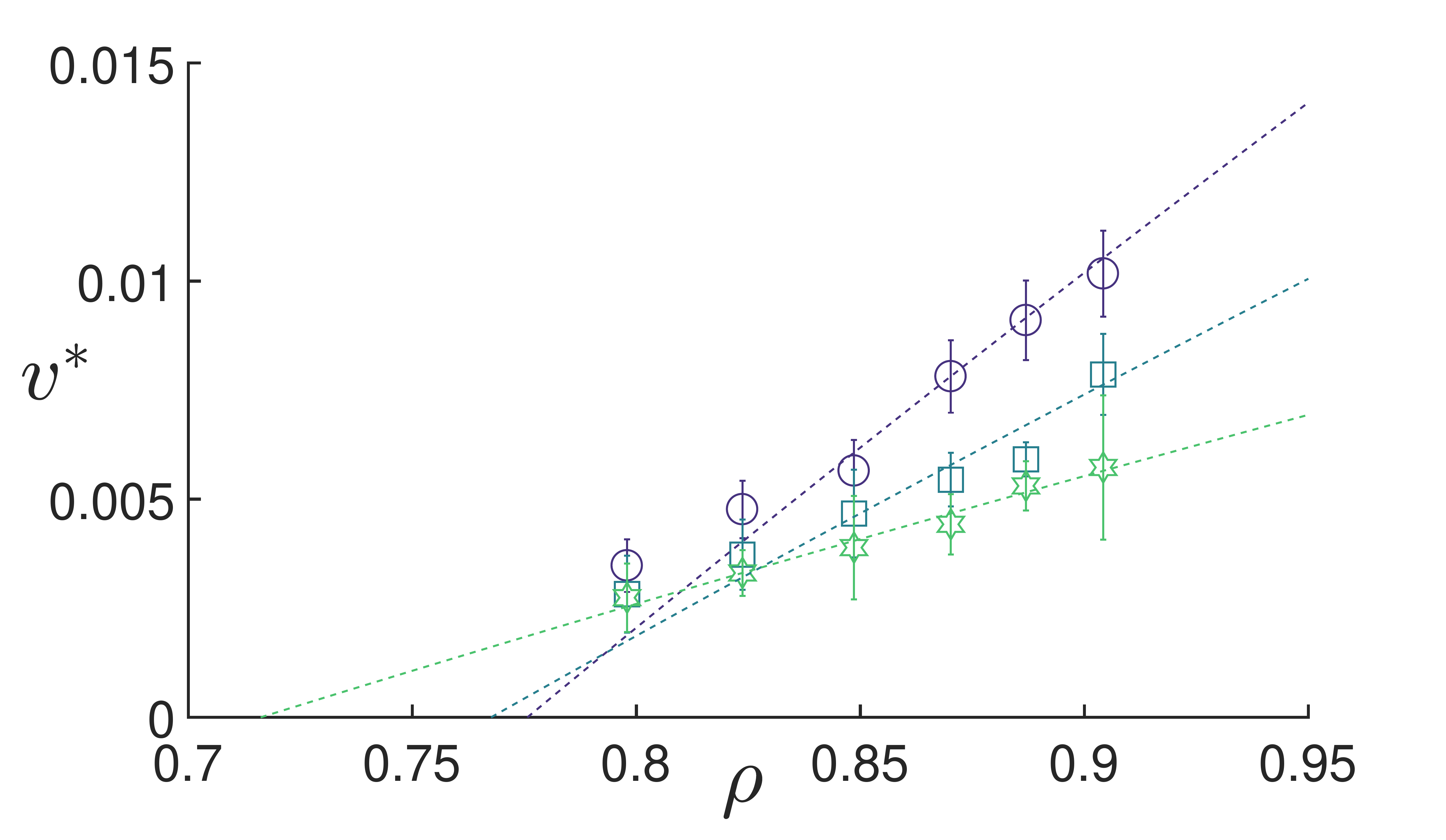}}}%
}

\selectlanguage{english}%
\caption{\textbf{Signatures of Jamming. (a}): snapshot of cell interface in
jammed state, colored lines trace cell center of mass over $t=8\cdot10^{4}$,
$\left(\textrm{active velocity }v_{A}=0.0075,\ \textrm{degree of confluence }\rho=87\%,\,\textrm{all normal cells}\right)$.
\textbf{(b)}: Mean square displacement over time for a range of increasing
active velocities shows initial ballistic motion $t\ll\tau=10^{4}$.
Late times show jammed behavior becoming diffusive as $v_{A}$ is
increased $\left(\rho=89\%,\,\textrm{all normal cells}\right)$. \textbf{(c)}:
Sixfold Bond correlations, $g_{6}(r)$ for a series of simulations
$\left(\rho=87\%,\,\textrm{all normal cells}\right)$ plotted on a
log-log scale. The dashed line is a guide to the eye at the predicted
KT algebraic decay $\eta=\frac{1}{4}$. \textbf{(d): }Effective diffusion
constant decreases linearly with $v_{A}$ before reaching a noise
floor $2\cdot10^{-3}$ $\left(\rho=89\%,\,\textrm{all normal cells}\right)$.
A best fit line (shown in green) is used to estimate the jamming velocity
$v^{*}$ (shown in red).\textbf{ (e)}: Diffusion with fixed \textbf{$v_{A}$
}with\textbf{ }varying\textbf{ }concentration for two parameter sets:
blue ($v_{A}=0.006$, $50\%$ mixing), and green ($v_{A}=0.003$,
100\% soft). Both are in good agreement with single-parameter fits
(dashed lines) of Eq. \ref{eq:Dvsrho}. \textbf{(f): }Jamming transition
velocity\textbf{ }as a function of concentration. Transition is consistent
with linear mixing. Fit lines are single parameter fits according
to Eq. \ref{eq:v vs rho}. \label{fig:v* scaling}\label{fig:Signatures-of-Jamming.}}
\end{figure*}
 The model can be used at any degree of confluence, or concentration,
$\rho=\frac{N\pi R_{0}^{2}}{L^{2}},$ where $N$ is the number of
cells and $L$ is the length of the simulation box. This does not
take into account the distance between adjacent interfaces which is
proportional to $\lambda$, and hence it under-represents slightly
the actual degree of confluence. Cell stiffness was increased three-fold
from our previous study, $\gamma=1.35$ for 'soft' cells and $\gamma=3.75$
for 'hard' cells, while $v_{A}$ was varied throughout. The rest of
the simulation parameters were unchanged, $\lambda=7,\ R_{0}=\lambda^{2},\ \mu=0.5,\ \tau=10^{4},\ \epsilon=10^{3}$
and time integration step $dt=0.1$, all given in a.u.. Each simulation
consists of $288$ cells in a square box with periodic boundary conditions.
Simulations were run for at least $t=800,000$ after an equilibration
of $t=80,000$ from an initial hexagonal configuration. We expect
that our results extend beyond the exact details of the model as the
long-time behavior has been shown to be independent of model details
in a dense liquid near the glass transition \cite{Gleim1998}.

We begin by demonstrating that our system can reach the jammed state.
Under the right conditions, the active velocity is unable to allow
a cell to squeeze through its neighbors leading to dynamical arrest.
Fig. \ref{fig:signatures-of-Jamming:snapshot} shows a snapshot of
cell interfaces and their center of mass movements over $t=8\cdot10^{4}$,
showing most cells do not exchange neighbors. Of course, that alone
is not sufficient evidence of jamming. As our system is active rather
than thermal, we can cross the jamming point by varying $v_{A}$,
the active motor strength. Keeping other parameters constant, Fig.
\ref{fig:signatures-of-Jamming:MSD} depicts the mean square displacement
(MSD) over time as a function of simulation time. For short times
$t\ll\tau$, all simulations show ballistic diffusion where $\text{MSD\ \ensuremath{\propto}}\ v_{A}^{2}t^{2}$.
By $t\gg\tau$ however, the behavior depends on $v_{A}.$ Large active
velocity (shades of green) have the expected diffusive behavior, $MSD\ \propto\ t$,
and the magnitude decreases with $v_{A}$. Eventually $v_{A}$ drops
below a threshold and MSD becomes sub-linear (shades of red) indicating
possible dynamical arrest due to caging effects. 
\begin{figure*}
\selectlanguage{american}%
\centering%
\makebox[0.45\paperwidth]{%
\subfigure[]{\label{fig:transition-diffusion}\foreignlanguage{english}{\includegraphics[width=3.5in]{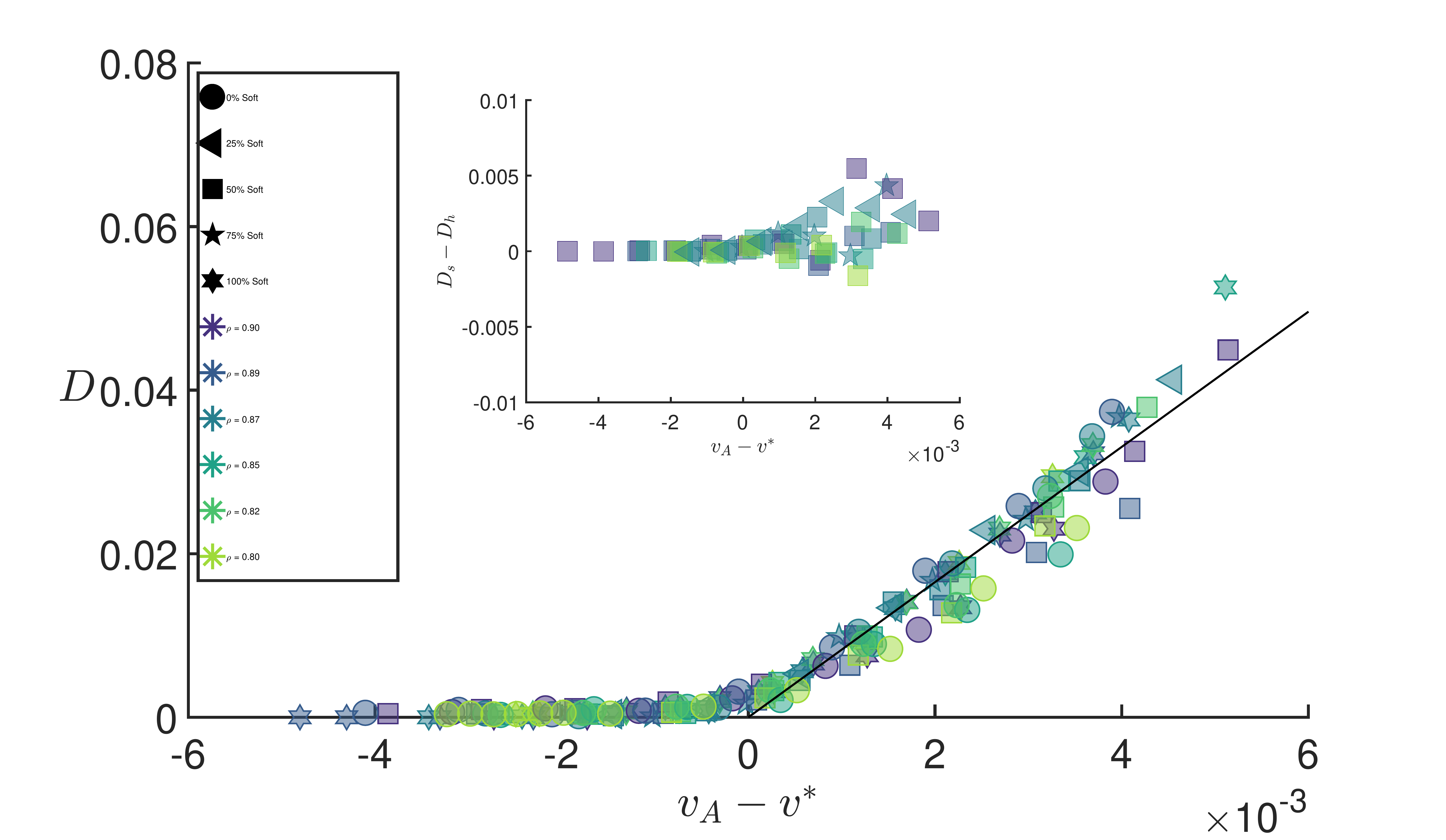}}}%
}\hspace{3in}\centering%
\makebox[0.45\paperwidth]{%
\subfigure[]{\foreignlanguage{english}{\includegraphics[width=3.5in]{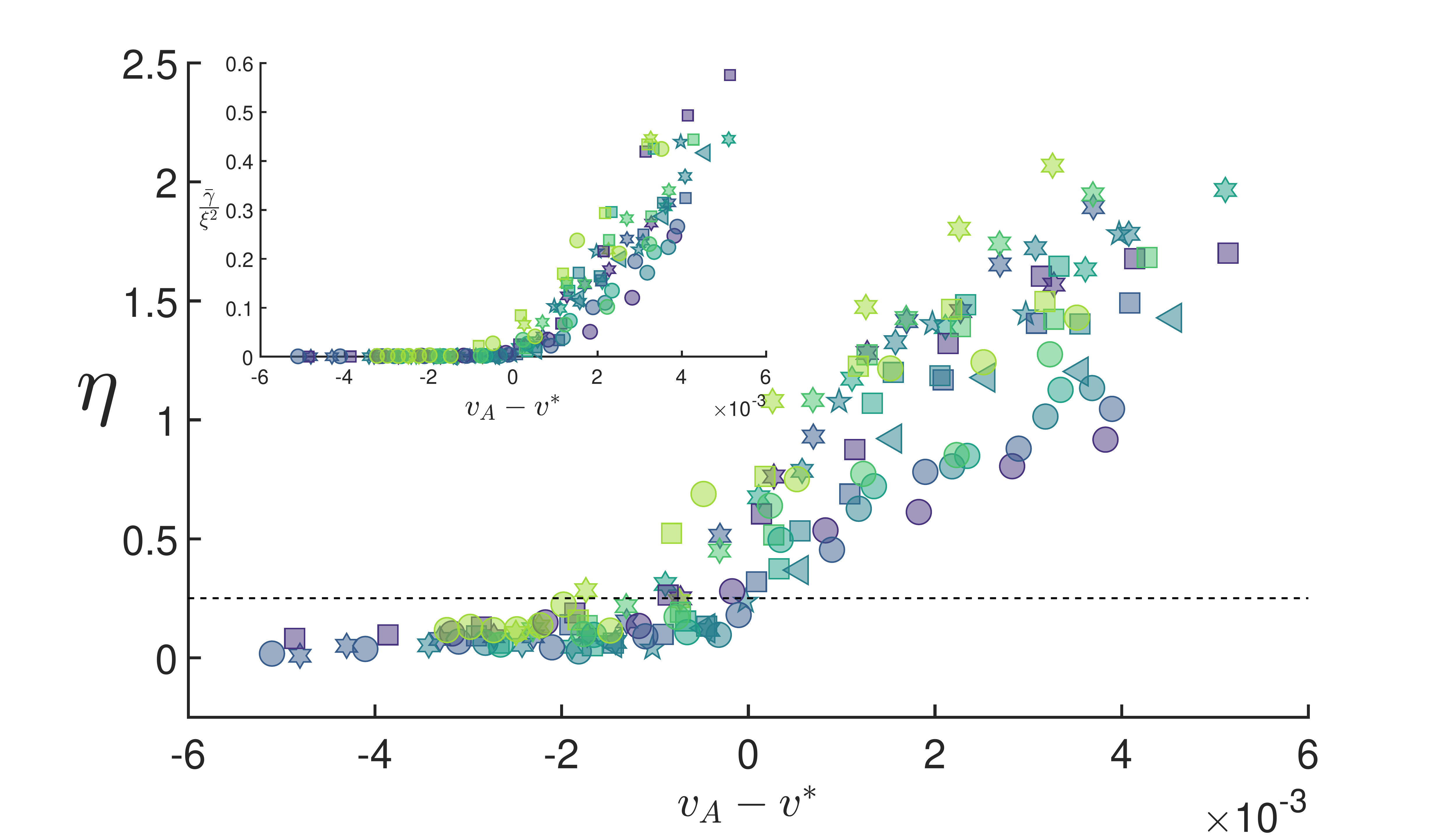}}\label{fig:transition-eta}}%
}\\
\centering%
\makebox[0.3\paperheight]{%
\subfigure[]{\label{fig:transition-qhard}\foreignlanguage{english}{\includegraphics[width=2in]{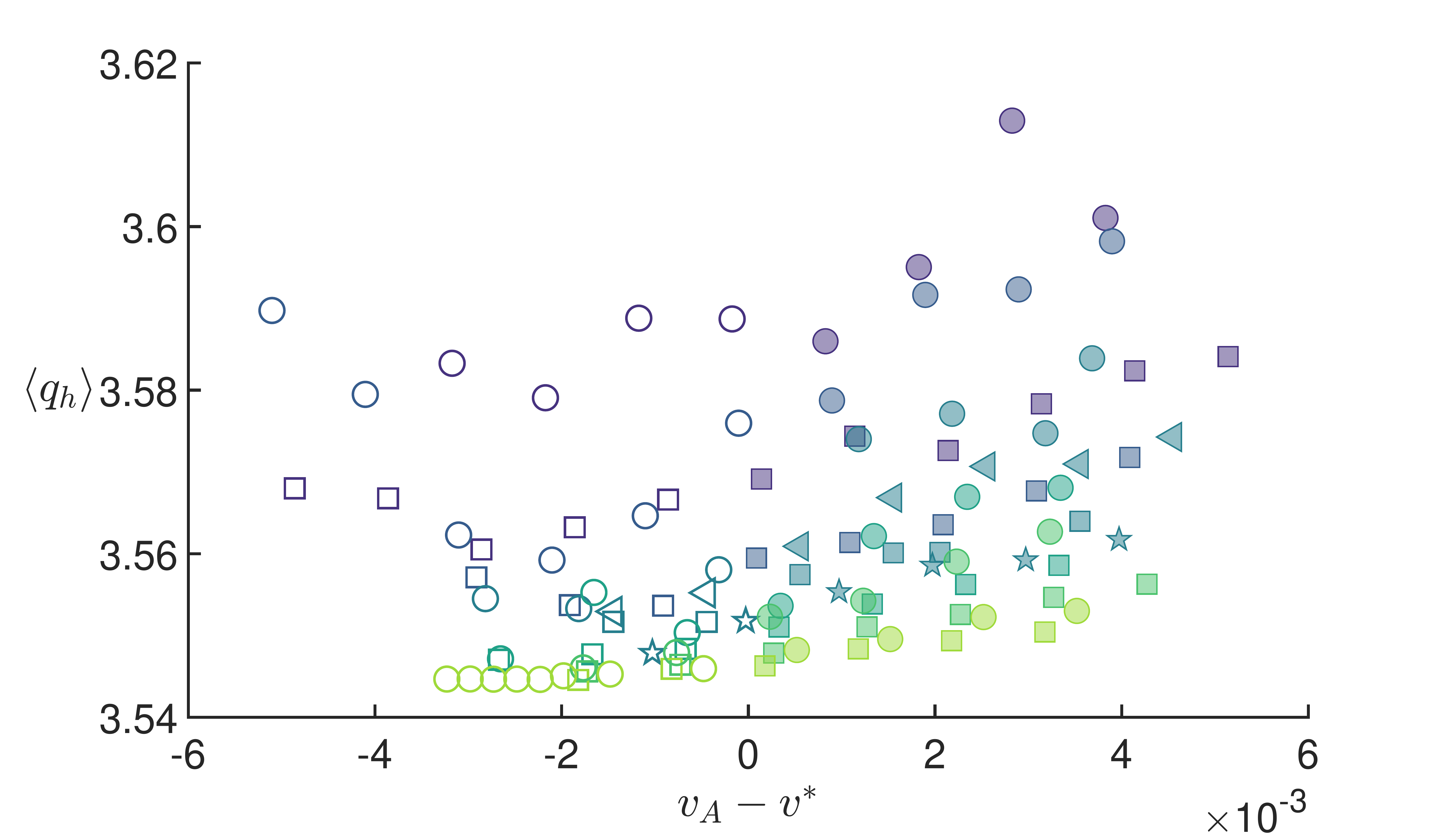}}}%
}\hspace{2in}\centering%
\makebox[0.3\paperwidth]{%
\subfigure[]{\foreignlanguage{english}{\includegraphics[width=2in]{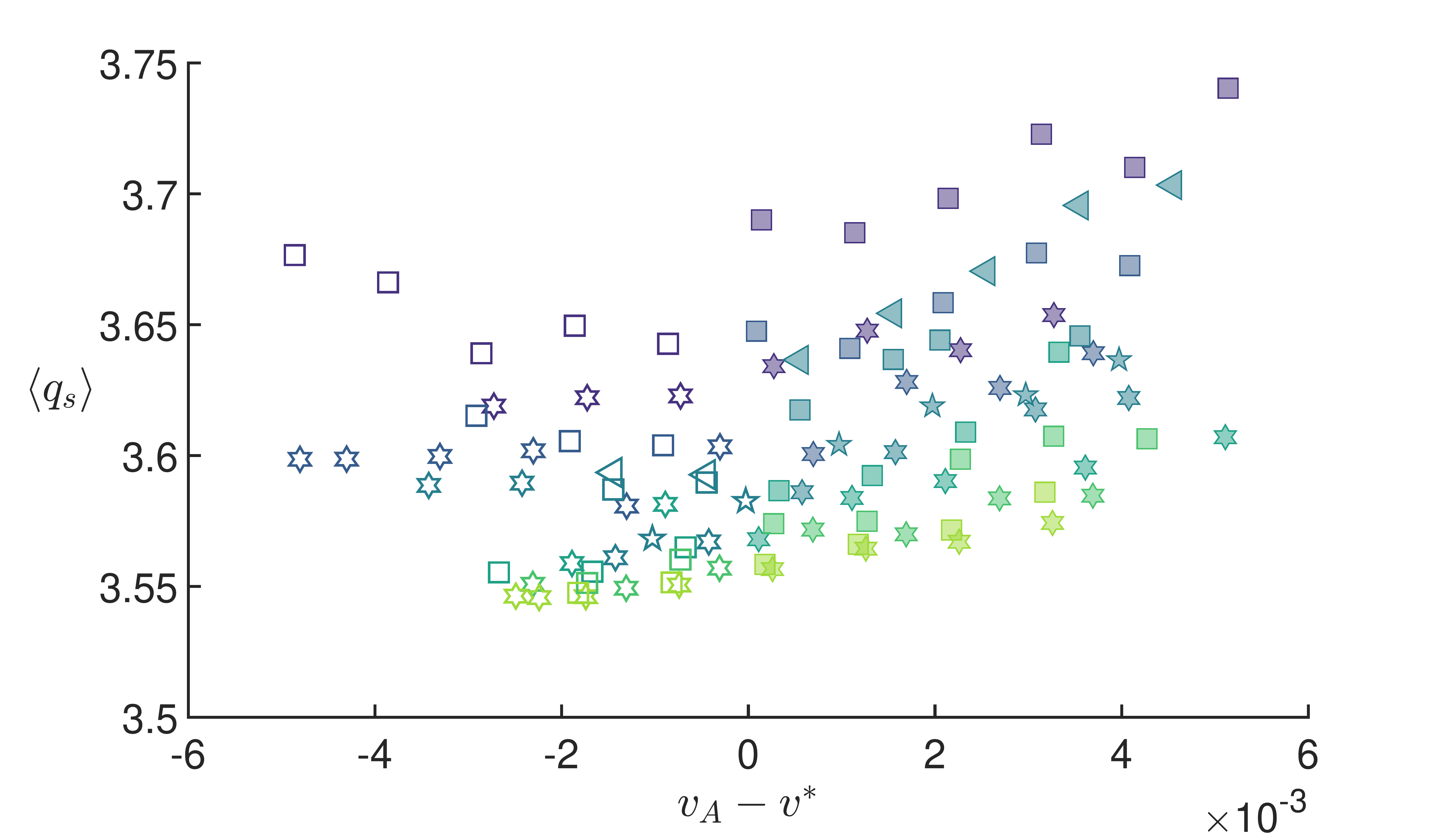}}\label{fig:transition-qsoft}}%
}\hspace{2in}\centering%
\makebox[0.3\paperwidth]{%
\subfigure[]{\foreignlanguage{english}{\includegraphics[width=2in]{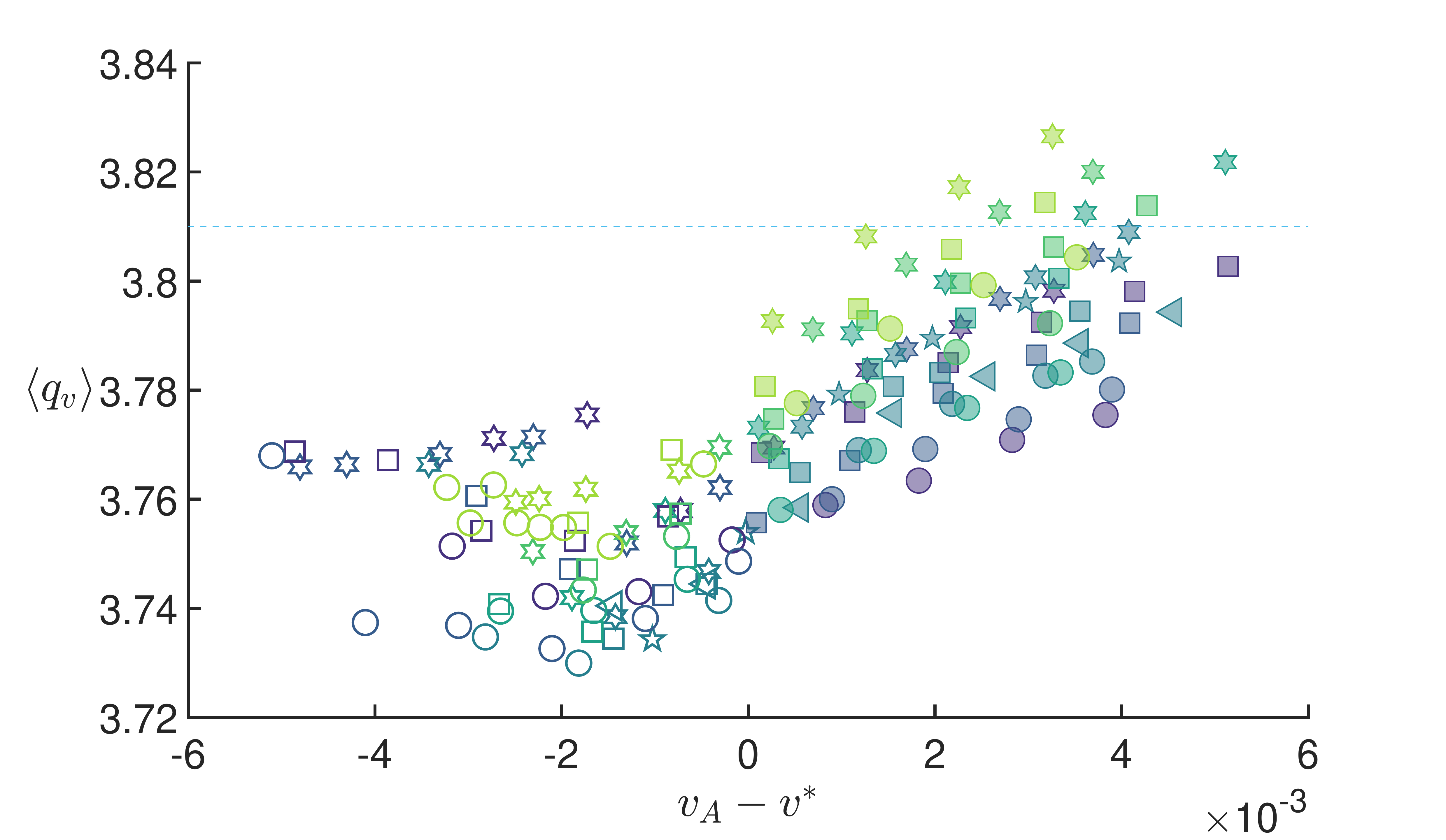}}\label{fig:transition-qV}}%
}

\selectlanguage{english}%
\caption{\textbf{Behavior across the jamming transition. }Symbol shape and
color denote mixing ratio and concentration, respectively, as shown
in legend top left. \textbf{(a): }The effective diffusion shows universal
behavior above the jamming transition. Solid line is a single parameter
line of best-fit. Inset: the difference between diffusion of soft
and hard cells within the same simulation. \textbf{(b): $\eta$ }(algebraic)
fit to $g_{6}$, consistent within error that the transition occurs
at the value predicted from KT theory, $\eta=\frac{1}{4}$ (shown
dashed line). Inset: Exponential correlation of the liquid state $\xi$
appears linear and universal upon rescaling.\label{fig:jamming compilation}\label{fig:DvsV compilation}
Average dimensionless shape index for hard cells \textbf{(c)}, soft
cells \textbf{(d)} and Voronoi tessellation \textbf{(e)}. For both
soft and hard cells, shape index generally increases after unjamming,
but transition value varies with simulation parameters. Using Cell
center positions from our simulations, the Voronoi tessellation shows
a more consistent value, though different from previous prediction
$\left\langle q\right\rangle _{SPP}=3.81$, shown as dashed line.
For all 3 figures, results in the jammed regime are shown with unfilled
symbols as they are sensitive to equilibration procedure. }
\end{figure*}
A more direct signature of a transition from a liquid to a jammed
state is in the falloff of the bond correlations of the sixfold orientation
order parameter \cite{Halperin1978} $\Psi_{6}\left(\mathbf{r}\right)=\sum_{j,k}e^{6i\theta_{jk}}$,
where the sum is over every neighbor $k$ within $2.5R_{0}$ of cell
j, and $\theta_{jk}\left(\mathbf{r}\right)$ is the angle between
them relative to the $x-axis$. A solid exhibits long range orientational
order, and the bond-orientational correlation function $g_{6}\left(r\right)=\left\langle \psi_{6}\left(\mathbf{r}\right)\psi_{6}^{*}\left(\mathbf{0}\right)\right\rangle $
approaches a nonzero constant for large $r$, whereas in a liquid
it decays exponentially with a characteristic correlation length $\xi$,
$g_{6}\left(r\right)\sim e^{-r/\xi}$. However, in between these may
lie an additional hexatic phase, where bond orientation decays algebraically,
$g_{6}\left(r\right)\propto r^{-\eta\left(T\right)}$ . The Kosterlitz-Thouless
theory also predicts the hexatic-liquid transition occurs when $\eta=\frac{1}{4}$
(and implicitly, the solid-hexatic transition is at $\eta=0$) \cite{Murray1992}.
Figure \ref{fig:signatures-of-Jamming:g6} shows the bond orientational
correlation function for a series of simulations. Low $v_{A}$ are
below the hexatic liquid transition shown as the dashed line. $\eta$
grows with increasing motor strength, crossing into the liquid phase
where the exponential fall-off would be faster than linear on this
log-log plot. Given our system size, it is not possible to discern
whether there exists an additional solid-hexatic phase transition.

We determine the transition point by examining the average effective
self-diffusion constant, $D$, as a function of $v_{A}$ in Fig. \ref{fig:signatures-of-Jamming:D_vs_v}.
It is computed using the Green-Kubo relation $D\left(t\right)=\frac{1}{2}\int_{0}^{t}\mathbf{\left\langle v\left(\mathrm{t}\mathrm{'}\right)\cdot v\left(\mathrm{0}\right)\right\rangle \mathrm{dt'}}$
and averaging over late times. We see $D$ decrease linearly with
$v_{A}$, until it vanishes below a noise floor of $D<2\cdot10^{-3}$.
Using a linear best fit we label the point at which diffusion vanishes
as $v^{*}$. We will show that this coincides with other signatures
of a jamming transition. See supplemental information for simulations
videos below $(v_{A}=0.008)$, very near $(v_{A}=0.009)$ and above
$(v_{A}=0.010)$ the jamming transition. Note these include the equilibration
time, which is discarded from all statistics.

We use a similar idea to the Turnbull free volume model to obtain
the effective diffusion for a thermal liquid near jamming \cite{Turnbull1970}:
consider a single cell with velocity $v_{A}$ moving towards a fixed
channel of width $d<2R+\lambda$. For the cell to squeeze through,
the dominant term of Eq. \ref{eq:drdt total} that must be overcome
is due to the curvature of the cell. Hence, there exists a minimum
velocity needed to fit through, which is given by $v^{*}\propto\frac{\lambda}{d}$.
If $v_{A}\leq v^{*}$, the cell will not be able to squeeze into the
channel and become jammed. Higher motor velocities will be able to
squeeze through and have net velocity $v_{eff}=v_{A}-v^{*}$, and
indeed, we see this in simulations of an isolated cells and immovable
square channel. Given this effective velocity, and that the system
is still diffusive, we obtain $D\propto\left(v-v^{*}\right)$. As
we will show, this model is also effective at predicting the transition
velocity as well as diffusion in the dilute regime.

Another method for detecting the transition expands on a result from
our previous studies \cite{Bresler2018} by investigating the effective
diffusion for a fixed $v_{A}$ as concentration is varied. Fig. \ref{fig:signatures-of-Jamming:D_vs_rho}
shows the result remains linear, but now we must account for jamming,
and 
\begin{equation}
D\left(\rho\right)=Max\left\{ D_{0}\left(1-\frac{\rho}{\rho^{*}}\right),0\right\} ,\label{eq:Dvsrho}
\end{equation}
 where $D_{0}=\frac{1}{2}v_{A}^{2}\tau$ is the diffusion of an isolated
cell, and $\rho^{*}$ is the concentration at which jamming occurs,
i.e. $v_{A}=v^{*}$. The single parameter lines of best-fit show good
agreement and suggest a method for estimating the onset of jamming
purely from the diffusion of unjammed cells in a more dilute concentration.
This differs from previous results of self-propelled soft disks \cite{Fily2012},
we expect due to details in the self-propulsion model.

We performed simulations over a range of concentrations ($0.92\%-0.80\%$)
and mixing ratios of soft to hard cells ($0,25,50,75\,\&\,100\%)$.
For each parameter set, $v_{A}$ was varied and a linear fit of $D\text{\ensuremath{\left(v_{A}\right)}}$
for values above the noise floor were used to estimate the transition
point, where $D\left(v^{*}\right)=0$. The results are summarized
in Figure \ref{fig:signatures-of-Jamming:v*_vs_rho}. Each series
represents different mixing ratios of soft and hard cells. As expected,
both higher concentration and harder cells lead to transition at higher
velocity. This reaffirms some of the known behavior of the jamming
phase diagram \cite{Donev2004}. Using the same simple model of the
cell squeezing through a channel, and the free volume between adjacent
hexagonal cells, we obtain
\begin{equation}
v^{*}\left(\rho\right)=\bar{\gamma}\frac{2\cdot3^{\frac{1}{4}}}{\sqrt{\pi}(R+\lambda/2)}\left(\sqrt{\rho}-\sqrt{\rho_{J}}\right),\label{eq:v vs rho}
\end{equation}
where $\bar{\gamma}$ is the average elasticity of all cells, and
$\rho_{J}$ is the concentration of point \textit{J} where jamming
occurs at zero velocity. This single parameter fit is shown as a dashed
line and is in good agreement with our results, particularly for the
harder cells and mixed cells, and gives similar estimates for $\rho_{J}$.
This is within error of previous results \cite{Donev2004,Olsson2011}
for soft repulsive discs, $\rho_{J}=0.843$, since our definition
of $\rho$ does not include the interface width. The soft cells, as
well as the results at the lowest concentration, appear to have larger
systematic error, likely due to the smaller values of $v*$ approaching
our noise floor. 

Having determined $v^{*}$ for each series, we now examine behavior
across the jamming transition, $v_{A}-v^{*}$ in Fig. \ref{fig:jamming compilation}.
In all plots, symbol color denotes concentration, while the symbol
shape denotes hard to soft cell mixing ratio, as shown in the legend.
Figure \ref{fig:transition-diffusion} shows a compilation for the
effective diffusion constant. $D\text{\ensuremath{\left(v_{A}-v^{*}\right)}}$
increases with a universal slope above the transition, as seen by
the line of best fit. Once again, this is consistent with the prediction
of our cell squeezing model. Furthermore, the best-fit slope $D\left(v-v^{*}\right)=8.264\left(v-v^{*}\right)$,
is surprisingly close to half the Turnbull coefficient \cite{Turnbull1961},
$2R/6\approx16.33$. In the case where both soft and hard cells are
present, the average value is shown. The inset shows the difference
in diffusion constant between soft and normal cells of the same simulation,
where the y-axis is $10$ times smaller than the main figure in order
to highlight the small differences. Although there might be a slight
bias for softer cells to have higher diffusion than normal cells in
the same simulation at large $v_{A}-v^{*}$, the cells appear to unjam
simultaneously as there is no discernible difference near $v_{A}-v^{*}\approx0$. 

Fig. \ref{fig:transition-eta} shows a similar compilation for $\eta$.
Here we find similar but non-universal scaling. Our results appear
consistent with the KT prediction $\eta\left(v-v^{*}=0\right)=\frac{1}{4}$,
though due to systematic error in determining $v^{*}$ and $\eta$,
we can not concretely rule out other results such as $\eta=0.385$
which has been suggested in other work \cite{Mazars2013}. As $\eta$
is poorly defined in the liquid state, the inset shows that the above
$v_{A}-v^{*}$ the liquid correlation length $\xi$ combined with
the average elasticity appears to be universal, as $\bar{\gamma}/\xi^{2}\propto v_{A}-v^{*}$. 

Bi \textit{et al.} showed \cite{Bi2016} that using a self-propelled
Voronoi, the average dimensionless shape index, $\left\langle q\right\rangle =\left\langle P\right\rangle /\sqrt{\left\langle A\right\rangle }$
(where $P$ is the cell perimeter) was a universal order parameter.
They found that at and below the transition exists a universal value
$\left\langle q\right\rangle _{SPP}=3.81$ , which increases above
the transition. We see that the same does not hold for this model.
The two panels show the average cell shape for hard (Fig. \ref{fig:transition-qhard})
and soft (Fig. \ref{fig:transition-qsoft}) cells. Results below the
jamming transitions are sensitive to equilibration procedure hence
they are shown with hollow symbols. The figures show that while the
shape index is increasing above the transition, the value varies significantly
between soft and hard cells, as well as with concentration and mixing
ratio, and hence is not a unique order parameter. 

Of course, the self-propelled Voronoi model differs from our sharp
interface in several ways, most significantly simulations are always
performed at full confluence $\rho=1$, cells have limited deformation,
and all cells had the same elasticity. We make a more direct comparison
by computing the Voronoi tessellation of the cell center of mass from
our simulation, similar to that used to analyze the local structure
and statistics of experimental cell data \cite{Rieser2016}. As shown
in Fig. \ref{fig:transition-qV}, the Voronoi shape index does appear
to be a good order parameter for the transition. But our value at
the transition $q_{V}^{*}\approx3.765$ is lower than the self-propelled
Voronoi model value ($q_{SPP}^{*}\approx3.81$, shown as dashed line).
Note that for $v_{A}-v^{*}<0$ the shape index appears to go below
the value at the transition for some simulations, in contrast to the
self-propelled Voronoi model. We suggest that although the Voronoi
shape index may be a useful order parameter, the transition value
appears to be model dependent, and hence may vary between cell lines
as well.

In conclusion, we have demonstrated that our elastic cell model reproduces
both liquid and hexatic phase. For each parameter set, we estimate
the velocity at the onset of jamming $v^{*}$, and our results are
consistent with the hallmarks of jamming. Using a simple free volume
argument along with simulation results, we have also shown the following
relations: an exact relation for the transition velocity, shown in
Eq. \ref{eq:v vs rho}. That the effective diffusion is zero in the
hexatic phase and has universal linear behavior with increasing $v_{A}$,
and that that diffusion for a fixed $v_{A}$ below the jamming concentration
depends on the jamming value. The hexatic-liquid transition appears
consistent with KT theory $\eta=\frac{1}{4}$, and in the liquid phase
$\bar{\gamma}/\xi^{2}\propto v_{A}-v^{*}$. Finally, we showed the
shape index $q$ of the actual cells $\left(q_{S},q_{H}\right)$ at
the transition varies with system parameters. Using the shape index
of the Voronoi tessellation, the transition appears constant at all
concentrations and mixing ratios, $q_{V}^{*}\approx3.765$, though
this is lower than the previously predicted universal value.

We gratefully acknowledged The Natural Sciences and Engineering Research
Council of Canada and the \textit{Fonds qu\'eb\'ecois de la recherche
sur la nature et les technologies} for funding this research, as well
as \textit{Calcul Qu\'ebec }and Compute Canada for providing computing
facilities.

\bibliographystyle{aipauth4-1}
\bibliography{Automatically_Imported}

\begin{thebibliography}{34}%
\makeatletter
\providecommand \@ifxundefined [1]{%
 \@ifx{#1\undefined}
}%
\providecommand \@ifnum [1]{%
 \ifnum #1\expandafter \@firstoftwo
 \else \expandafter \@secondoftwo
 \fi
}%
\providecommand \@ifx [1]{%
 \ifx #1\expandafter \@firstoftwo
 \else \expandafter \@secondoftwo
 \fi
}%
\providecommand \natexlab [1]{#1}%
\providecommand \enquote  [1]{``#1''}%
\providecommand \bibnamefont  [1]{#1}%
\providecommand \bibfnamefont [1]{#1}%
\providecommand \citenamefont [1]{#1}%
\providecommand \href@noop [0]{\@secondoftwo}%
\providecommand \href [0]{\begingroup \@sanitize@url \@href}%
\providecommand \@href[1]{\@@startlink{#1}\@@href}%
\providecommand \@@href[1]{\endgroup#1\@@endlink}%
\providecommand \@sanitize@url [0]{\catcode `\\12\catcode `\$12\catcode
  `\&12\catcode `\#12\catcode `\^12\catcode `\_12\catcode `\%12\relax}%
\providecommand \@@startlink[1]{}%
\providecommand \@@endlink[0]{}%
\providecommand \url  [0]{\begingroup\@sanitize@url \@url }%
\providecommand \@url [1]{\endgroup\@href {#1}{\urlprefix }}%
\providecommand \urlprefix  [0]{URL }%
\providecommand \Eprint [0]{\href }%
\providecommand \doibase [0]{http://dx.doi.org/}%
\providecommand \selectlanguage [0]{\@gobble}%
\providecommand \bibinfo  [0]{\@secondoftwo}%
\providecommand \bibfield  [0]{\@secondoftwo}%
\providecommand \translation [1]{[#1]}%
\providecommand \BibitemOpen [0]{}%
\providecommand \bibitemStop [0]{}%
\providecommand \bibitemNoStop [0]{.\EOS\space}%
\providecommand \EOS [0]{\spacefactor3000\relax}%
\providecommand \BibitemShut  [1]{\csname bibitem#1\endcsname}%
\let\auto@bib@innerbib\@empty
\bibitem [{\citenamefont {Angelini}\ \emph {et~al.}(2011)\citenamefont
  {Angelini}, \citenamefont {Hannezo}, \citenamefont {Trepat}, \citenamefont
  {Marquez}, \citenamefont {Fredberg},\ and\ \citenamefont
  {Weitz}}]{Angelini2011}%
  \BibitemOpen
  \bibfield  {author} {\bibinfo {author} {\bibnamefont {Angelini},
  \bibfnamefont {T.~E.}}, \bibinfo {author} {\bibnamefont {Hannezo},
  \bibfnamefont {E.}}, \bibinfo {author} {\bibnamefont {Trepat}, \bibfnamefont
  {X.}}, \bibinfo {author} {\bibnamefont {Marquez}, \bibfnamefont {M.}},
  \bibinfo {author} {\bibnamefont {Fredberg}, \bibfnamefont {J.~J.}}, \ and\
  \bibinfo {author} {\bibnamefont {Weitz}, \bibfnamefont {D.~A.}},\ }\href
  {\doibase 10.1073/pnas.1010059108} {\bibfield  {journal} {\bibinfo  {journal}
  {Proceedings of the National Academy of Sciences}\ }\textbf {\bibinfo
  {volume} {108}},\ \bibinfo {pages} {4714} (\bibinfo {year}
  {2011})}\BibitemShut {NoStop}%
\bibitem [{\citenamefont {Bernard}\ and\ \citenamefont
  {Krauth}(2011)}]{Bernard2011}%
  \BibitemOpen
  \bibfield  {author} {\bibinfo {author} {\bibnamefont {Bernard}, \bibfnamefont
  {E.~P.}}\ and\ \bibinfo {author} {\bibnamefont {Krauth}, \bibfnamefont
  {W.}},\ }\href {\doibase 10.1103/PhysRevLett.107.155704} {\bibfield
  {journal} {\bibinfo  {journal} {Physical Review Letters}\ }\textbf {\bibinfo
  {volume} {107}},\ \bibinfo {pages} {1} (\bibinfo {year} {2011})},\ \Eprint
  {http://arxiv.org/abs/1102.4094} {arXiv:1102.4094} \BibitemShut {NoStop}%
\bibitem [{\citenamefont {Berthier}\ and\ \citenamefont
  {Kurchan}(2013)}]{Berthier2013}%
  \BibitemOpen
  \bibfield  {author} {\bibinfo {author} {\bibnamefont {Berthier},
  \bibfnamefont {L.}}\ and\ \bibinfo {author} {\bibnamefont {Kurchan},
  \bibfnamefont {J.}},\ }\href {\doibase 10.1038/nphys2592} {\bibfield
  {journal} {\bibinfo  {journal} {Nature Physics}\ }\textbf {\bibinfo {volume}
  {9}},\ \bibinfo {pages} {310} (\bibinfo {year} {2013})},\ \Eprint
  {http://arxiv.org/abs/1302.4868} {arXiv:1302.4868} \BibitemShut {NoStop}%
\bibitem [{\citenamefont {Berthier}\ and\ \citenamefont
  {Tarjus}(2009)}]{Berthier2009}%
  \BibitemOpen
  \bibfield  {author} {\bibinfo {author} {\bibnamefont {Berthier},
  \bibfnamefont {L.}}\ and\ \bibinfo {author} {\bibnamefont {Tarjus},
  \bibfnamefont {G.}},\ }\href {\doibase 10.1103/PhysRevLett.103.170601}
  {\bibfield  {journal} {\bibinfo  {journal} {Physical Review Letters}\
  }\textbf {\bibinfo {volume} {103}},\ \bibinfo {pages} {23} (\bibinfo {year}
  {2009})},\ \Eprint {http://arxiv.org/abs/0907.2343} {arXiv:0907.2343}
  \BibitemShut {NoStop}%
\bibitem [{\citenamefont {Bi}\ \emph {et~al.}(2014)\citenamefont {Bi},
  \citenamefont {Lopez}, \citenamefont {Schwarz},\ and\ \citenamefont
  {Manning}}]{Bi2015}%
  \BibitemOpen
  \bibfield  {author} {\bibinfo {author} {\bibnamefont {Bi}, \bibfnamefont
  {D.}}, \bibinfo {author} {\bibnamefont {Lopez}, \bibfnamefont {J.~H.}},
  \bibinfo {author} {\bibnamefont {Schwarz}, \bibfnamefont {J.~M.}}, \ and\
  \bibinfo {author} {\bibnamefont {Manning}, \bibfnamefont {M.~L.}},\ }\href
  {\doibase 10.1038/nphys3471} {\bibfield  {journal} {\bibinfo  {journal}
  {Nature Physics}\ }\textbf {\bibinfo {volume} {11}},\ \bibinfo {pages} {1074}
  (\bibinfo {year} {2014})},\ \Eprint {http://arxiv.org/abs/1409.0593}
  {arXiv:1409.0593} \BibitemShut {NoStop}%
\bibitem [{\citenamefont {Bi}\ \emph {et~al.}(2016)\citenamefont {Bi},
  \citenamefont {Yang}, \citenamefont {Marchetti},\ and\ \citenamefont
  {Manning}}]{Bi2016}%
  \BibitemOpen
  \bibfield  {author} {\bibinfo {author} {\bibnamefont {Bi}, \bibfnamefont
  {D.}}, \bibinfo {author} {\bibnamefont {Yang}, \bibfnamefont {X.}}, \bibinfo
  {author} {\bibnamefont {Marchetti}, \bibfnamefont {M.~C.}}, \ and\ \bibinfo
  {author} {\bibnamefont {Manning}, \bibfnamefont {M.~L.}},\ }\href {\doibase
  10.1103/PhysRevX.6.021011} {\bibfield  {journal} {\bibinfo  {journal}
  {Physical Review X}\ }\textbf {\bibinfo {volume} {6}},\ \bibinfo {pages} {1}
  (\bibinfo {year} {2016})},\ \Eprint {http://arxiv.org/abs/1509.06578}
  {arXiv:1509.06578} \BibitemShut {NoStop}%
\bibitem [{\citenamefont {Bresler}, \citenamefont {Palmieri},\ and\
  \citenamefont {Grant}(2018)}]{Bresler2018}%
  \BibitemOpen
  \bibfield  {author} {\bibinfo {author} {\bibnamefont {Bresler}, \bibfnamefont
  {Y.}}, \bibinfo {author} {\bibnamefont {Palmieri}, \bibfnamefont {B.}}, \
  and\ \bibinfo {author} {\bibnamefont {Grant}, \bibfnamefont {M.}},\ }\href
  {http://arxiv.org/abs/1807.07836} {\ ,\ \bibinfo {pages} {1} (\bibinfo {year}
  {2018})},\ \Eprint {http://arxiv.org/abs/cond-mat.soft/1807.07836}
  {arXiv:cond-mat.soft/1807.07836} \BibitemShut {NoStop}%
\bibitem [{\citenamefont {Brookes}(2017)}]{Brookes2017}%
  \BibitemOpen
  \bibfield  {author} {\bibinfo {author} {\bibnamefont {Brookes}, \bibfnamefont
  {N.~H.}},\ }\href@noop {} {\  (\bibinfo {year} {2017})}\BibitemShut {NoStop}%
\bibitem [{\citenamefont {Chepizhko}\ \emph {et~al.}(2018)\citenamefont
  {Chepizhko}, \citenamefont {Lionetti}, \citenamefont {Malinverno},
  \citenamefont {Scita}, \citenamefont {Zapperi},\ and\ \citenamefont {{La
  Porta}}}]{Chepizhko2018}%
  \BibitemOpen
  \bibfield  {author} {\bibinfo {author} {\bibnamefont {Chepizhko},
  \bibfnamefont {O.}}, \bibinfo {author} {\bibnamefont {Lionetti},
  \bibfnamefont {M.~C.}}, \bibinfo {author} {\bibnamefont {Malinverno},
  \bibfnamefont {C.}}, \bibinfo {author} {\bibnamefont {Scita}, \bibfnamefont
  {G.}}, \bibinfo {author} {\bibnamefont {Zapperi}, \bibfnamefont {S.}}, \ and\
  \bibinfo {author} {\bibnamefont {{La Porta}}, \bibfnamefont {C.~A.~M.}},\
  }\href {http://arxiv.org/abs/1802.07488} {\  (\bibinfo {year} {2018})},\
  \Eprint {http://arxiv.org/abs/1802.07488} {arXiv:1802.07488} \BibitemShut
  {NoStop}%
\bibitem [{\citenamefont {Delarue}\ \emph {et~al.}(2016)\citenamefont
  {Delarue}, \citenamefont {Hartung}, \citenamefont {Schreck}, \citenamefont
  {Gniewek}, \citenamefont {Hu}, \citenamefont {Herminghaus},\ and\
  \citenamefont {Hallatschek}}]{Delarue2016}%
  \BibitemOpen
  \bibfield  {author} {\bibinfo {author} {\bibnamefont {Delarue}, \bibfnamefont
  {M.}}, \bibinfo {author} {\bibnamefont {Hartung}, \bibfnamefont {J.}},
  \bibinfo {author} {\bibnamefont {Schreck}, \bibfnamefont {C.}}, \bibinfo
  {author} {\bibnamefont {Gniewek}, \bibfnamefont {P.}}, \bibinfo {author}
  {\bibnamefont {Hu}, \bibfnamefont {L.}}, \bibinfo {author} {\bibnamefont
  {Herminghaus}, \bibfnamefont {S.}}, \ and\ \bibinfo {author} {\bibnamefont
  {Hallatschek}, \bibfnamefont {O.}},\ }\href {\doibase 10.1038/nphys3741}
  {\bibfield  {journal} {\bibinfo  {journal} {Nature Physics}\ }\textbf
  {\bibinfo {volume} {12}},\ \bibinfo {pages} {762} (\bibinfo {year}
  {2016})}\BibitemShut {NoStop}%
\bibitem [{\citenamefont {Fily}\ and\ \citenamefont
  {Marchetti}(2012)}]{Fily2012}%
  \BibitemOpen
  \bibfield  {author} {\bibinfo {author} {\bibnamefont {Fily}, \bibfnamefont
  {Y.}}\ and\ \bibinfo {author} {\bibnamefont {Marchetti}, \bibfnamefont
  {M.~C.}},\ }\href {\doibase 10.1103/PhysRevLett.108.235702} {\bibfield
  {journal} {\bibinfo  {journal} {Physical Review Letters}\ }\textbf {\bibinfo
  {volume} {108}},\ \bibinfo {pages} {1} (\bibinfo {year} {2012})},\ \Eprint
  {http://arxiv.org/abs/1201.4847} {arXiv:1201.4847} \BibitemShut {NoStop}%
\bibitem [{\citenamefont {Gleim}, \citenamefont {Kob},\ and\ \citenamefont
  {Binder}(1998)}]{Gleim1998}%
  \BibitemOpen
  \bibfield  {author} {\bibinfo {author} {\bibnamefont {Gleim}, \bibfnamefont
  {T.}}, \bibinfo {author} {\bibnamefont {Kob}, \bibfnamefont {W.}}, \ and\
  \bibinfo {author} {\bibnamefont {Binder}, \bibfnamefont {K.}},\ }\href
  {\doibase 10.1103/PhysRevLett.81.4404} {\bibfield  {journal} {\bibinfo
  {journal} {Physical Review Letters}\ }\textbf {\bibinfo {volume} {81}},\
  \bibinfo {pages} {4404} (\bibinfo {year} {1998})},\ \Eprint
  {http://arxiv.org/abs/9805200} {arXiv:9805200 [cond-mat]} \BibitemShut
  {NoStop}%
\bibitem [{\citenamefont {Haeger}\ \emph {et~al.}(2014)\citenamefont {Haeger},
  \citenamefont {Krause}, \citenamefont {Wolf},\ and\ \citenamefont
  {Friedl}}]{Haeger2014a}%
  \BibitemOpen
  \bibfield  {author} {\bibinfo {author} {\bibnamefont {Haeger}, \bibfnamefont
  {A.}}, \bibinfo {author} {\bibnamefont {Krause}, \bibfnamefont {M.}},
  \bibinfo {author} {\bibnamefont {Wolf}, \bibfnamefont {K.}}, \ and\ \bibinfo
  {author} {\bibnamefont {Friedl}, \bibfnamefont {P.}},\ }\href {\doibase
  10.1016/j.bbagen.2014.03.020} {\bibfield  {journal} {\bibinfo  {journal}
  {Biochimica et Biophysica Acta - General Subjects}\ }\textbf {\bibinfo
  {volume} {1840}},\ \bibinfo {pages} {2386} (\bibinfo {year}
  {2014})}\BibitemShut {NoStop}%
\bibitem [{\citenamefont {Halperin}\ and\ \citenamefont
  {Nelson}(1978)}]{Halperin1978}%
  \BibitemOpen
  \bibfield  {author} {\bibinfo {author} {\bibnamefont {Halperin},
  \bibfnamefont {B.~I.}}\ and\ \bibinfo {author} {\bibnamefont {Nelson},
  \bibfnamefont {D.~R.}},\ }\href {\doibase 10.1103/PhysRevLett.41.121}
  {\bibfield  {journal} {\bibinfo  {journal} {Physical Review Letters}\
  }\textbf {\bibinfo {volume} {41}},\ \bibinfo {pages} {121} (\bibinfo {year}
  {1978})}\BibitemShut {NoStop}%
\bibitem [{\citenamefont {Henkes}, \citenamefont {Fily},\ and\ \citenamefont
  {Marchetti}(2011)}]{Henkes2011}%
  \BibitemOpen
  \bibfield  {author} {\bibinfo {author} {\bibnamefont {Henkes}, \bibfnamefont
  {S.}}, \bibinfo {author} {\bibnamefont {Fily}, \bibfnamefont {Y.}}, \ and\
  \bibinfo {author} {\bibnamefont {Marchetti}, \bibfnamefont {M.~C.}},\ }\href
  {\doibase 10.1103/PhysRevE.84.040301} {\bibfield  {journal} {\bibinfo
  {journal} {Physical Review E - Statistical, Nonlinear, and Soft Matter
  Physics}\ }\textbf {\bibinfo {volume} {84}},\ \bibinfo {pages} {84} (\bibinfo
  {year} {2011})},\ \Eprint {http://arxiv.org/abs/1107.4072} {arXiv:1107.4072}
  \BibitemShut {NoStop}%
\bibitem [{\citenamefont {Kosterlitz}\ and\ \citenamefont
  {Thouless}(1973)}]{Kosterlitz1973}%
  \BibitemOpen
  \bibfield  {author} {\bibinfo {author} {\bibnamefont {Kosterlitz},
  \bibfnamefont {J.~M.}}\ and\ \bibinfo {author} {\bibnamefont {Thouless},
  \bibfnamefont {D.~J.}},\ }\href {\doibase 10.1088/0022-3719/6/7/010}
  {\bibfield  {journal} {\bibinfo  {journal} {Journal of Physics C: Solid State
  Physics}\ }\textbf {\bibinfo {volume} {6}},\ \bibinfo {pages} {1181}
  (\bibinfo {year} {1973})}\BibitemShut {NoStop}%
\bibitem [{\citenamefont {Langer}(2014)}]{Langer2014}%
  \BibitemOpen
  \bibfield  {author} {\bibinfo {author} {\bibnamefont {Langer}, \bibfnamefont
  {J.~S.}},\ }\href {\doibase 10.1088/0034-4885/77/4/042501} {\bibfield
  {journal} {\bibinfo  {journal} {Reports on Progress in Physics}\ }\textbf
  {\bibinfo {volume} {77}} (\bibinfo {year} {2014}),\
  10.1088/0034-4885/77/4/042501},\ \Eprint {http://arxiv.org/abs/1308.6544}
  {arXiv:1308.6544} \BibitemShut {NoStop}%
\bibitem [{\citenamefont {Madhikar}\ \emph {et~al.}(2018)\citenamefont
  {Madhikar}, \citenamefont {{\AA}str{\"{o}}m}, \citenamefont {Westerholm},\
  and\ \citenamefont {Karttunen}}]{Madhikar2018}%
  \BibitemOpen
  \bibfield  {author} {\bibinfo {author} {\bibnamefont {Madhikar},
  \bibfnamefont {P.}}, \bibinfo {author} {\bibnamefont {{\AA}str{\"{o}}m},
  \bibfnamefont {J.}}, \bibinfo {author} {\bibnamefont {Westerholm},
  \bibfnamefont {J.}}, \ and\ \bibinfo {author} {\bibnamefont {Karttunen},
  \bibfnamefont {M.}},\ }\href {\doibase 10.1016/j.cpc.2018.05.024} {\bibfield
  {journal} {\bibinfo  {journal} {Computer Physics Communications}\ } (\bibinfo
  {year} {2018}),\ 10.1016/j.cpc.2018.05.024}\BibitemShut {NoStop}%
\bibitem [{\citenamefont {Mazars}(2013)}]{Mazars2013}%
  \BibitemOpen
  \bibfield  {author} {\bibinfo {author} {\bibnamefont {Mazars}, \bibfnamefont
  {M.}},\ }\href {http://arxiv.org/abs/1301.1571} {\bibfield  {journal}
  {\bibinfo  {journal} {Arxiv preprint}\ ,\ \bibinfo {pages} {14}} (\bibinfo
  {year} {2013})},\ \Eprint {http://arxiv.org/abs/1301.1571} {arXiv:1301.1571}
  \BibitemShut {NoStop}%
\bibitem [{\citenamefont {Murray}(1992)}]{Murray1992}%
  \BibitemOpen
  \bibfield  {author} {\bibinfo {author} {\bibnamefont {Murray}, \bibfnamefont
  {C.~A.}},\ }in\ \href {\doibase 10.1007/978-1-4612-2812-7_4} {\emph {\bibinfo
  {booktitle} {Bond-Orientational Order in Condensed Matter Systems. Partially
  Ordered Systems. Springer, New York, NY}}}\ (\bibinfo {year} {1992})\ pp.\
  \bibinfo {pages} {137--215}\BibitemShut {NoStop}%
\bibitem [{\citenamefont {Nogucci}(2018)}]{Nogucci2018}%
  \BibitemOpen
  \bibfield  {author} {\bibinfo {author} {\bibnamefont {Nogucci}, \bibfnamefont
  {H.}},\ }\href {http://arxiv.org/abs/1802.09149} {\  (\bibinfo {year}
  {2018})},\ \Eprint {http://arxiv.org/abs/1802.09149} {arXiv:1802.09149}
  \BibitemShut {NoStop}%
\bibitem [{\citenamefont {O'Hern}\ \emph {et~al.}(2003)\citenamefont {O'Hern},
  \citenamefont {Silbert}, \citenamefont {Liu},\ and\ \citenamefont
  {Nagel}}]{Donev2004}%
  \BibitemOpen
  \bibfield  {author} {\bibinfo {author} {\bibnamefont {O'Hern}, \bibfnamefont
  {C.~S.}}, \bibinfo {author} {\bibnamefont {Silbert}, \bibfnamefont {L.~E.}},
  \bibinfo {author} {\bibnamefont {Liu}, \bibfnamefont {A.~J.}}, \ and\
  \bibinfo {author} {\bibnamefont {Nagel}, \bibfnamefont {S.~R.}},\ }\href
  {\doibase 10.1103/PhysRevE.68.011306} {\bibfield  {journal} {\bibinfo
  {journal} {Physical Review E}\ }\textbf {\bibinfo {volume} {68}},\ \bibinfo
  {pages} {011306} (\bibinfo {year} {2003})},\ \Eprint
  {http://arxiv.org/abs/0304421v1} {arXiv:0304421v1 [arXiv:cond-mat]}
  \BibitemShut {NoStop}%
\bibitem [{\citenamefont {Olsson}\ and\ \citenamefont
  {Teitel}(2011)}]{Olsson2011}%
  \BibitemOpen
  \bibfield  {author} {\bibinfo {author} {\bibnamefont {Olsson}, \bibfnamefont
  {P.}}\ and\ \bibinfo {author} {\bibnamefont {Teitel}, \bibfnamefont {S.}},\
  }\href {\doibase 10.1103/PhysRevE.83.030302} {\bibfield  {journal} {\bibinfo
  {journal} {Physical Review E - Statistical, Nonlinear, and Soft Matter
  Physics}\ }\textbf {\bibinfo {volume} {83}},\ \bibinfo {pages} {2} (\bibinfo
  {year} {2011})},\ \Eprint {http://arxiv.org/abs/1010.5885} {arXiv:1010.5885}
  \BibitemShut {NoStop}%
\bibitem [{\citenamefont {Oswald}\ \emph {et~al.}(2017)\citenamefont {Oswald},
  \citenamefont {Grosser}, \citenamefont {Smith},\ and\ \citenamefont
  {K{\"{a}}s}}]{Oswald2017}%
  \BibitemOpen
  \bibfield  {author} {\bibinfo {author} {\bibnamefont {Oswald}, \bibfnamefont
  {L.}}, \bibinfo {author} {\bibnamefont {Grosser}, \bibfnamefont {S.}},
  \bibinfo {author} {\bibnamefont {Smith}, \bibfnamefont {D.~M.}}, \ and\
  \bibinfo {author} {\bibnamefont {K{\"{a}}s}, \bibfnamefont {J.~A.}},\ }\href
  {\doibase 10.1088/1361-6463/aa8e83} {\bibfield  {journal} {\bibinfo
  {journal} {Journal of Physics D: Applied Physics}\ }\textbf {\bibinfo
  {volume} {50}} (\bibinfo {year} {2017}),\
  10.1088/1361-6463/aa8e83}\BibitemShut {NoStop}%
\bibitem [{\citenamefont {Palmieri}\ \emph {et~al.}(2015)\citenamefont
  {Palmieri}, \citenamefont {Bresler}, \citenamefont {Wirtz},\ and\
  \citenamefont {Grant}}]{Palmieri2015}%
  \BibitemOpen
  \bibfield  {author} {\bibinfo {author} {\bibnamefont {Palmieri},
  \bibfnamefont {B.}}, \bibinfo {author} {\bibnamefont {Bresler}, \bibfnamefont
  {Y.}}, \bibinfo {author} {\bibnamefont {Wirtz}, \bibfnamefont {D.}}, \ and\
  \bibinfo {author} {\bibnamefont {Grant}, \bibfnamefont {M.}},\ }\href
  {\doibase 10.1038/srep11745} {\bibfield  {journal} {\bibinfo  {journal}
  {Scientific Reports}\ }\textbf {\bibinfo {volume} {5}},\ \bibinfo {pages}
  {11745} (\bibinfo {year} {2015})}\BibitemShut {NoStop}%
\bibitem [{\citenamefont {Parisi}\ and\ \citenamefont
  {Zamponi}(2010)}]{Parisi2010}%
  \BibitemOpen
  \bibfield  {author} {\bibinfo {author} {\bibnamefont {Parisi}, \bibfnamefont
  {G.}}\ and\ \bibinfo {author} {\bibnamefont {Zamponi}, \bibfnamefont {F.}},\
  }\href {\doibase 10.1103/RevModPhys.82.789} {\bibfield  {journal} {\bibinfo
  {journal} {Reviews of Modern Physics}\ }\textbf {\bibinfo {volume} {82}},\
  \bibinfo {pages} {789} (\bibinfo {year} {2010})},\ \Eprint
  {http://arxiv.org/abs/0802.2180} {arXiv:0802.2180} \BibitemShut {NoStop}%
\bibitem [{\citenamefont {Peng}\ \emph {et~al.}(2010)\citenamefont {Peng},
  \citenamefont {Wang}, \citenamefont {Alsayed}, \citenamefont {Yodh},\ and\
  \citenamefont {Han}}]{Peng2010}%
  \BibitemOpen
  \bibfield  {author} {\bibinfo {author} {\bibnamefont {Peng}, \bibfnamefont
  {Y.}}, \bibinfo {author} {\bibnamefont {Wang}, \bibfnamefont {Z.}}, \bibinfo
  {author} {\bibnamefont {Alsayed}, \bibfnamefont {A.~M.}}, \bibinfo {author}
  {\bibnamefont {Yodh}, \bibfnamefont {A.~G.}}, \ and\ \bibinfo {author}
  {\bibnamefont {Han}, \bibfnamefont {Y.}},\ }\href {\doibase
  10.1103/PhysRevLett.104.205703} {\bibfield  {journal} {\bibinfo  {journal}
  {Physical Review Letters}\ }\textbf {\bibinfo {volume} {104}},\ \bibinfo
  {pages} {205703} (\bibinfo {year} {2010})}\BibitemShut {NoStop}%
\bibitem [{\citenamefont {Prestipino}, \citenamefont {Saija},\ and\
  \citenamefont {Giaquinta}(2011)}]{Prestipino2011}%
  \BibitemOpen
  \bibfield  {author} {\bibinfo {author} {\bibnamefont {Prestipino},
  \bibfnamefont {S.}}, \bibinfo {author} {\bibnamefont {Saija}, \bibfnamefont
  {F.}}, \ and\ \bibinfo {author} {\bibnamefont {Giaquinta}, \bibfnamefont
  {P.~V.}},\ }\href {\doibase 10.1103/PhysRevLett.106.235701} {\bibfield
  {journal} {\bibinfo  {journal} {Physical Review Letters}\ }\textbf {\bibinfo
  {volume} {106}},\ \bibinfo {pages} {235701} (\bibinfo {year}
  {2011})}\BibitemShut {NoStop}%
\bibitem [{\citenamefont {Rieser}\ \emph {et~al.}(2016)\citenamefont {Rieser},
  \citenamefont {Goodrich}, \citenamefont {Liu},\ and\ \citenamefont
  {Durian}}]{Rieser2016}%
  \BibitemOpen
  \bibfield  {author} {\bibinfo {author} {\bibnamefont {Rieser}, \bibfnamefont
  {J.~M.}}, \bibinfo {author} {\bibnamefont {Goodrich}, \bibfnamefont {C.~P.}},
  \bibinfo {author} {\bibnamefont {Liu}, \bibfnamefont {A.~J.}}, \ and\
  \bibinfo {author} {\bibnamefont {Durian}, \bibfnamefont {D.~J.}},\ }\href
  {\doibase 10.1103/PhysRevLett.116.088001} {\bibfield  {journal} {\bibinfo
  {journal} {Physical Review Letters}\ }\textbf {\bibinfo {volume} {116}},\
  \bibinfo {pages} {1} (\bibinfo {year} {2016})},\ \Eprint
  {http://arxiv.org/abs/1509.05496} {arXiv:1509.05496} \BibitemShut {NoStop}%
\bibitem [{\citenamefont {Sadati}\ \emph {et~al.}(2013)\citenamefont {Sadati},
  \citenamefont {{Taheri Qazvini}}, \citenamefont {Krishnan}, \citenamefont
  {Park},\ and\ \citenamefont {Fredberg}}]{Sadati2013}%
  \BibitemOpen
  \bibfield  {author} {\bibinfo {author} {\bibnamefont {Sadati}, \bibfnamefont
  {M.}}, \bibinfo {author} {\bibnamefont {{Taheri Qazvini}}, \bibfnamefont
  {N.}}, \bibinfo {author} {\bibnamefont {Krishnan}, \bibfnamefont {R.}},
  \bibinfo {author} {\bibnamefont {Park}, \bibfnamefont {C.~Y.}}, \ and\
  \bibinfo {author} {\bibnamefont {Fredberg}, \bibfnamefont {J.~J.}},\ }\href
  {\doibase 10.1016/j.diff.2013.02.005} {\bibfield  {journal} {\bibinfo
  {journal} {Differentiation}\ }\textbf {\bibinfo {volume} {86}},\ \bibinfo
  {pages} {121} (\bibinfo {year} {2013})},\ \Eprint
  {http://arxiv.org/abs/NIHMS150003} {arXiv:NIHMS150003} \BibitemShut {NoStop}%
\bibitem [{\citenamefont {Sillescu}(1999)}]{Sillescu1999}%
  \BibitemOpen
  \bibfield  {author} {\bibinfo {author} {\bibnamefont {Sillescu},
  \bibfnamefont {H.}},\ }\href {\doibase 10.1016/S0022-3093(98)00831-X}
  {\bibfield  {journal} {\bibinfo  {journal} {Journal of Non-Crystalline
  Solids}\ }\textbf {\bibinfo {volume} {243}},\ \bibinfo {pages} {81} (\bibinfo
  {year} {1999})}\BibitemShut {NoStop}%
\bibitem [{\citenamefont {Steinhardt}, \citenamefont {Nelson},\ and\
  \citenamefont {Ronchetti}(1983)}]{Steinhardt1983}%
  \BibitemOpen
  \bibfield  {author} {\bibinfo {author} {\bibnamefont {Steinhardt},
  \bibfnamefont {P.~J.}}, \bibinfo {author} {\bibnamefont {Nelson},
  \bibfnamefont {D.~R.}}, \ and\ \bibinfo {author} {\bibnamefont {Ronchetti},
  \bibfnamefont {M.}},\ }\href {\doibase 10.1103/PhysRevB.28.784} {\bibfield
  {journal} {\bibinfo  {journal} {Physical Review B}\ }\textbf {\bibinfo
  {volume} {28}},\ \bibinfo {pages} {784} (\bibinfo {year} {1983})}\BibitemShut
  {NoStop}%
\bibitem [{\citenamefont {Turnbull}\ and\ \citenamefont
  {Cohen}(1961)}]{Turnbull1961}%
  \BibitemOpen
  \bibfield  {author} {\bibinfo {author} {\bibnamefont {Turnbull},
  \bibfnamefont {D.}}\ and\ \bibinfo {author} {\bibnamefont {Cohen},
  \bibfnamefont {M.~H.}},\ }\href {\doibase 10.1063/1.1731549} {\bibfield
  {journal} {\bibinfo  {journal} {The Journal of Chemical Physics}\ }\textbf
  {\bibinfo {volume} {34}},\ \bibinfo {pages} {120} (\bibinfo {year} {1961})},\
  \Eprint {http://arxiv.org/abs/arXiv:1011.1669v3} {arXiv:arXiv:1011.1669v3}
  \BibitemShut {NoStop}%
\bibitem [{\citenamefont {Turnbull}\ and\ \citenamefont
  {Cohen}(1970)}]{Turnbull1970}%
  \BibitemOpen
  \bibfield  {author} {\bibinfo {author} {\bibnamefont {Turnbull},
  \bibfnamefont {D.}}\ and\ \bibinfo {author} {\bibnamefont {Cohen},
  \bibfnamefont {M.~H.}},\ }\href {\doibase 10.1063/1.1673434} {\bibfield
  {journal} {\bibinfo  {journal} {The Journal of Chemical Physics}\ }\textbf
  {\bibinfo {volume} {52}},\ \bibinfo {pages} {3038} (\bibinfo {year}
  {1970})},\ \Eprint {http://arxiv.org/abs/arXiv:1011.1669v3}
  {arXiv:arXiv:1011.1669v3} \BibitemShut {NoStop}%
\end{thebibliography}%

\end{document}